\documentclass[twocolumn,showpacs,preprintnumbers,amsmath,amssymb,superscriptaddress,pra]{revtex4}
\usepackage{dcolumn}
\usepackage{bm}
\usepackage{graphicx}

\newcommand{\ket}[1]{\displaystyle{|#1\rangle}}
\newcommand{\bra}[1]{\displaystyle{\langle #1|}}

\newcommand{\sumkj}{\sum_{\mathbf{k}j}}

\newcommand{\sumkkjj}{\sum_{\mathbf{k}\mathbf{k'}jj'}}
\newcommand{\akj}{a_{\mathbf{k}j}}
\newcommand{\ackj}{a^{\dag}_{\mathbf{k}j}}
\newcommand{\akjp}{a_{\mathbf{k}'j'}}

\newcommand{\omegak}{\omega_{\mathbf{k}}}
\newcommand{\omegakp}{\omega_{\mathbf{k'}}}
\newcommand{\omegaz}{\omega_{0}}
\newcommand{\fkjrA}{\mathbf{f}_{\mathbf{k}j}(\mathbf{r}_A)}

\newcommand{\fkjR}{\mathbf{f}_{\mathbf{k}j}(\mathbf{R})}

\newcommand{\fkjr}{\mathbf{f}_{\mathbf{k}j}(\mathbf{r})}
\newcommand{\fkjpr}{\mathbf{f}_{\mathbf{k'}j'}(\mathbf{r})}

\newcommand{\fkj}{\mathbf{f}_{\mathbf{k}j}}

\newcommand{\Sp}{S_{+}}
\newcommand{\Sm}{S_{-}}
\newcommand{\Sz}{S_{z}}

\newcommand{\rv}{\mathbf{r}}

\newcommand{\ekj}{\hat{e}_{\mathbf{k}j}}
\newcommand{\ekjx}{(\hat{e}_{\mathbf{k}j})_{x}}
\newcommand{\ekjy}{(\hat{e}_{\mathbf{k}j})_{y}}
\newcommand{\ekjz}{(\hat{e}_{\mathbf{k}j})_{z}}

\def\bbm[#1]{\mbox{\boldmath $#1$}}

\DeclareMathOperator{\Si}{Si}
\DeclareMathOperator{\Ci}{Ci}

\begin{document}

\title{Dynamical Casimir-Polder force on a partially dressed atom near a conducting wall}
\author{Riccardo Messina}
\affiliation{SYRTE - Observatoire de Paris
61, avenue de l'Observatoire, F-75014 Paris, France}
\author{Ruggero Vasile}
\affiliation{Turku Centre for Quantum Physics, Department of Physics
and Astronomy, University of Turku, 20014 Turun Yliopisto, Finland}
\author{Roberto Passante}
\affiliation{Dipartimento di Scienze Fisiche e Astronomiche
dell'Universit\`{a} degli Studi di Palermo and CNSIM, Via Archirafi 36,
I-90123 Palermo, Italy}

\date{\today}

\begin{abstract}
We study the time evolution of the Casimir-Polder force acting on a
neutral atom in front of a perfectly conducting plate, when the
system starts its unitary evolution from a partially dressed state.
We solve the Heisenberg equations for both atomic and field quantum
operators, exploiting a series expansion with respect to the
electric charge and an iterative technique. After discussing the
behaviour of the time-dependent force on an initially
partially-dressed atom, we analyze a possible experimental scheme to
prepare the partially dressed state and the observability of this
new dynamical effect.
\end{abstract}


\maketitle

\section{Introduction}

According to quantum electrodynamics, the electric and magnetic
fields show unavoidable fluctuations around their average values,
even in the ground state of the field \cite{Milonni,CPP95}. This
feature gives rise to many physical phenomena such as the existence
of a force between a couple of electrically neutral but polarizable
objects. The existence of this kind of forces was first remarked by
Casimir in 1948 for two parallel, neutral and perfectly conducting
plates \cite{CasimirProcKonNederlAkadWet48}, and by Casimir and
Polder for a neutral atom in front of a plate as well as between two
neutral atoms \cite{CasimirPhysRev48}. The force between an atom and
a surface, which is the main topic of this paper, has been measured
with remarkable precision, notwithstanding the smallness of the
force, using different techniques: deflection of atomic beams sent
in proximity of surfaces \cite{SukenikPRL93}, reflection of cold
atoms \cite{LandraginPRL96,ShimizuPRL01,DruzhininaPRL03}. More
recently, Bose-Einstein condensates were exploited to obtain more
precise measurements of the atom-surface force, using both
reflection techniques \cite{PasquiniPRL04,PasquiniPRL06} and the
observation of center-of-mass oscillations of the condensate
\cite{AntezzaPRA04,HarberPRA05,AntezzaPRL06,ObrechtPRL07}.

The inclusion of dynamical (time-dependent) aspects in the system
can considerably change the physical nature of the observed
phenomena. When dealing with the dynamical Casimir-Polder effect, it
is worthwhile distinguishing two different possible realizations of
the dynamics. A first important situation to consider is the
mechanical movement of the bodies of the system: in this case the
emission of real photons can take place, having its dissipative
counterpart in a friction force acting on the moving objects. This
idea was first brought to attention in the pioneering works of Moore
\cite{MooreJMathPhys76} and Fulling and Davies
\cite{FullingProcRSocLondA76}, and then it paved the way to a
remarkable amount of theoretical work (see \cite{DodonovArxiv} and
references therein). It does not yet exist any experimental
observation of the emission of radiation by dynamical Casimir
effect, due to the very low rate of photon emission, but a promising
experiment is currently in progress in which the mechanical movement
is replaced by the periodical modulation of the optical properties
of one of the surfaces involved
\cite{DodonovArxiv,BraggioRevSciInstrum04,BraggioEurophysLett05}. On
the other hand, the expression \emph{dynamical Casimir-Polder force}
is also used in the discussion of the time dependence of the force
if the system undergoes a unitary evolution starting from a
non-equilibrium quantum state
\cite{PassantePhysLettA03,RizzutoPRA04}. For example, in
\cite{PassantePhysLettA03} the authors studied the time evolution of
the force between two neutral atoms starting from a partially
dressed state of the system. Such a state is an intermediate
configuration between the bare ground state of the system, which is
given by the tensor product of the atomic ground state and the
vacuum field state, and the physical, completely dressed, ground
state of the composite system. Although the papers
\cite{VasilePRA08,ShrestaPRA03} deal with the physical configuration
we are interested in, that is a neutral atom in front of a
conducting wall, the evolution is there studied starting from the
bare ground state of the system, which is an idealized configuration
hardly achievable in the laboratory.

In this paper we consider the evolution in time of the force between
an atom and a perfectly conducting infinite plate starting from a
partially dressed state, which is a much more realistic physical
situation. To this aim, we are going to exploit, in analogy with
\cite{VasilePRA08}, the method introduced by Power and
Thirunamachandran \cite{PowerPRA83B} for atoms in free space. It
consists in solving the Heisenberg equations of the atomic and field
operators in the Heisenberg picture by performing a series expansion
with respect to the coupling constant (the electric charge) and then
iteratively finding the solution (see \cite{PowerPRA83B} or
\cite{RizzutoPRA04} for more details). Then the time-dependent
atom-wall Casimir-Polder energy is obtained for a specific model of
a partially dressed atom, obtained by a rapid change of the atomic
transition frequency due to an external action on the atom such as
an external electric field
\cite{PassantePhysLettA96,PassantePhysLettA03}. Finally, we discuss
experimental realizability of the model considered and possibility
of observing the dynamical effects predicted by our results.

This paper is organized as follows. In Section \ref{Sec:2} we
introduce the multipolar coupling scheme for a two-level atom
interacting with the radiation field in the electric dipole
approximation and in the presence of a conducting wall. Then we
solve the Heisenberg equations for the photon creation and
annihilation operators up to the first order in the electric charge,
using an iterative technique, in order to obtain the Heisenberg
operator giving the time-dependent atom-wall interaction energy. The
solutions so obtained are valid for any initial state of the system.
In Section \ref{Sec:3} we discuss our choice of the initial state of
the system, that is a partially dressed atomic state, and evaluate
the time-dependent atom-wall interaction energy. Finally, in Sec.
\ref{Sec:4} we discuss in more detail our physical model, its
experimental realizability and the possible observation of the
predicted dynamical Casimir-Polder interaction.

\section{The Hamiltonian model}\label{Sec:2}

We consider a two-level atom interacting with the electromagnetic
radiation field in the presence of an infinite and perfectly
conducting wall. We let the mirror coincide with the plane $z=0$ and
we place the atom on its right side: the atomic position vector is
thus $\mathbf{r}_A\equiv(0,0,d)$, with $d>0$. We work in the
multipolar coupling scheme and within the electric dipole
approximation (see e.g. \cite{PowerPhilTransRoySocA59,PowerPRA83A}).
Thus the Hamiltonian describing our system reads
\begin{widetext}
\begin{equation}\begin{split}\label{IntHam}
H&=H_{0}+H_{I}\\
H_{0}&=\hbar\omegaz\Sz+\sumkj\hbar\omegak\ackj\akj\\
H_{I}&=-i\sqrt{\frac{2\pi\hbar
c}{V}}\sumkj\sqrt{k}(\bbm[\mu]\cdot\fkjrA)\bigl(\akj-\ackj\bigr)\left( \Sp + \Sm \right).\end{split}\end{equation}
\end{widetext}
In this expression the radiation field is described by the set of
bosonic annihilation and creation operators $\akj$ and $\ackj$,
associated with a photon of frequency $\omegak=ck$, while the matrix
element of the electric dipole moment operator $\bbm[\mu]$ and the
pseudospin operators $\Sp$, $\Sm$ and $\Sz$ are associated to the
atom, which has a transition frequency $\omega_0$ \cite{CPP95}.
Moreover $\fkj(\rv)$ are the field mode functions in the presence of
the wall, that in Eq. \eqref{IntHam} are evaluated at the atomic
position $\mathbf{r_A}$. Their expressions can be obtained from the
mode functions of a perfectly conducting cubical cavity of volume
$V=L^3$ with walls ($-L/2<x,\; y<L/2$, \;$0<z<L$)
\cite{Milonni,PowerPRA82}
\begin{widetext}
\begin{equation}\label{Funzmodoparete}\begin{split}
(\fkj(\rv))_{x}&=\sqrt{8}\ekjx\cos\Bigl[k_{x}\Bigl(x+\frac{L}2\Bigr)\Bigr]\sin\Bigl[k_{y}\Bigl(y+\frac L2\Bigr)\Bigr]\sin\left(k_z z\right)\\
(\fkj(\rv))_{y}&=\sqrt{8}\ekjy\sin\Bigl[k_{x}\Bigl(x+\frac{L}2\Bigr)\Bigr]\cos\Bigl[k_{y}\Bigl(y+\frac L2\Bigr)\Bigr]\sin\left(k_z z\right)\\
(\fkj(\rv))_{z}&=\sqrt{8}\ekjz\sin\Bigl[k_{x}\Bigl(x+\frac{L}2\Bigr)\Bigr]\sin\Bigl[k_{y}\Bigl(y+\frac L2\Bigr)\Bigr]\cos\left(k_ zz\right)
\end{split}\end{equation}
\end{widetext}
where $k_x=l\pi/L$,\; $k_y=m\pi/L$,\; $k_z=n\pi/L$
($l,m,n=0,1,2,\dots$) and $\ekj$ are polarization unit vectors. In
order to switch from the cavity to the wall at $z=0$, at the end of
the calculations one has to take the limit $L\rightarrow\infty$.

We are going to obtain all the information about the time evolution
of the atom-wall force by solving the Heisenberg equations of all the
atomic and field operators involved in our system. As anticipated
before, since it is not possible to solve exactly these equations
for our model, we shall use an iterative technique. As a starting
point we write the operators as a power series in the coupling
constant, that as an example for the annihilation operator takes the
form
\begin{equation}\label{powerseries}\akj(t)=\akj^{(0)}(t)+\akj^{(1)}(t)+\akj^{(2)}(t)+\dots\end{equation}
where the contribution $\akj^{(i)}(t)$ is proportional to the $i$-th
power of the electric charge. For our purposes we need the
expressions of both field and atomic operators up to the the first
order only. The result is already reported in \cite{VasilePRA08} and
it has the form
\begin{widetext}
\begin{equation}\label{SolOper}\begin{split}
\akj^{(0)}(t)&=e^{-i\omegak t}\akj\qquad S_\pm^{(0)}(t)=e^{\pm i\omega_0 t}S_\pm\\
\akj^{(1)}(t)&=e^{-i\omegak t}\sqrt{\frac{2\pi ck}{\hbar
V}}\bigl(\bbm[\mu]\cdot\fkjrA\bigl) \bigl[\Sp F(\omegak+\omegaz,t)+\Sm
F(\omegak-\omegaz,t)\bigl]\\
S^{(1)}_\pm(t)&=\mp2\Sz e^{\pm i\omega_0t}\sumkj\sqrt{\frac{2\pi ck}{\hbar V}}(\bbm[\mu]\cdot\fkjR)\Bigl(\akj F^*(\omegak\pm\omega_0,t)-\akj^\dag F(\omegak\mp\omega_0,t)\Bigr).\\
\end{split}\end{equation}
\end{widetext}
where we have introduced the auxiliary function
\begin{equation}\label{FunzioniF}
F(\omega,t)=\int_{0}^{t}e^{i\omega t'}dt'=\frac{e^{i\omega t}-1}{i\omega}.
\end{equation}
All operators appearing in the RHS of Eq. \eqref{SolOper} without
explicit time dependence are evaluated at $t=0$, and thus coincide
with their counterpart in the Schr\"{o}dinger picture. While the
zeroth-order terms correspond to the absence of interaction and then
to the free evolution given by $H_0$, the first-order terms couple
the atomic and field operators. We wish to stress here the main
advantage of solving the Heisenberg equations for the operators
involved in the system: since in the Heisenberg picture only the
operators evolve in time whilst the quantum state of the system
remains constant, when calculating the time evolution of any average
value the choice of the initial state can be performed just as a
final step.

\section{Choice of the initial state and interaction energy}\label{Sec:3}

Our aim is to calculate the time-dependent atom-wall interaction
energy, in particular for a partially dressed initial state. Using
the same method as in \cite{VasilePRA08}, valid in a quasi-static
approach at the second order, we shall calculate this quantity by
taking half of the average value on the initial state of the
interaction Hamiltonian $H_I^{(2)}(t)$ in the Heisenberg
representation. Then we have
\begin{equation}\label{Average}\Delta E^{(2)}(t)=\frac 12 \bra{\psi(0)}H^{(2)}_I(t)\ket{\psi(0)}\end{equation}
where $\ket{\psi(0)}$ is the initial state of the atom-field system.
The explicit expression of $H_I(t)$ up to the second order is easily
deduced from \eqref{IntHam} and \eqref{SolOper} (only atomic and
field operators up to the first order are necessary), and it is
given by
\begin{equation}\begin{split}&H_I^{(2)}(t)=-\frac{2\pi ic}{V}\sumkj k\bigl(\bbm[\mu]\cdot\fkjr\bigr)^2[\Sp e^{i\omegaz t}+\text{h.c.}]\{\Sp\\
&\,\times[e^{-i\omegak t}F(\omegaz+\omegak,t)-e^{i\omegak t}F^*(\omegak-\omegaz,t)]-\text{h.c.}\}\\
&\,+\frac{4\pi ic}{V}\Sz\sumkkjj\sqrt{kk'}(\bbm[\mu]\cdot\fkjr)(\bbm[\mu]\cdot\fkjpr)\\
&\,\times\{\akjp[e^{i\omegaz t}F^*(\omegaz+\omegakp,t)-e^{-i\omegaz t}\\
&\,\times F^*(\omegakp-\omegaz,t)]+\text{h.c.}\}[\akj e^{-i\omegak
t}-\text{h.c.}]\\\end{split}\end{equation}

Now we must choose a specific initial quantum state to be used in
\eqref{Average}. In \cite{VasilePRA08} the bare ground state was
considered as initial state. This state is the eigenstate of $H_0$
having minimum energy: it accounts to a switching off of the
interaction between atom and field and thus, although being a useful
idealization, it is difficult to imagine an experimental scheme for
generating such a state (but it can give important hints on the
behaviour of more realistic systems). On the contrary, the
completely dressed ground state of $H$ can be obtained by using
stationary perturbation theory, and its expression up to the first
order is given by
\begin{equation}\label{StatoCorretto}\begin{split}\ket{0_d}&=\ket{0}+\ket{1}\\
\ket{1}&=-i\sqrt{\frac{2\pi}{\hbar
V}}\sumkj\frac{\sqrt{k}(\bbm[\mu]\cdot\mathbf{f}_{\mathbf{k}j}(\mathbf{r}))}{k+k_0}\ket{1_{\mathbf{k}j},\uparrow}\end{split}\end{equation}
written as a sum of the bare ground state $\ket{0}$ and a sum of
one-photon states gathered in $\ket{1}$. Up to the first order in
the coupling constant, the state \eqref{StatoCorretto} does not
undergo any time evolution. This expression clearly depends on the
atomic transition frequency $ck_0$. This consideration is the basis
of our proposal for the preparation of a partially-dressed state: we
assume our atom initially to have a transition frequency $\omega_0'$
and to be in its completely dressed ground state (given by
\eqref{StatoCorretto} with $k_0'$ in place of $k_0$), and then to
produce at $t=0$ an abrupt change of its transition frequency from
$\omega_0'$ to a new frequency $\omega_0$. In the next Section we
shall discuss in more detail how this rapid change of the atom's
transition frequency could be obtained. From the physical point of
view, our hypothesis is that this change is so rapid that the
quantum state immediately after $t=0$ remains the same as before.
Thus this state will be taken as initial state of the unitary
evolution for $t>0$, given by the Hamiltonian \eqref{IntHam} with
$\omega_0$ as the value of the atom's transition frequency; this
state is subjected to a time evolution because it is not an
eigenstate of the \emph{new} Hamiltonian at $t>0$, which is that for
an atom with the new transition frequency $\omega_0$. A partially
dressed state is so obtained
\cite{PassantePhysLettA96,PassantePhysLettA03}.

We can now calculate three different average values of the
interaction energy. The first is obtained starting from the
completely dressed state of the system, given by
\eqref{StatoCorretto}: this state is a stationary state, and then we
simply recover the well known result for the static atom-wall force.
If, on the contrary, we consider the evolution from the bare ground
state $\ket{0\!\downarrow}$, we indeed observe a time evolution of
the atom-wall force, as obtained in \cite{VasilePRA08}. Finally, we
can choose as initial state $\ket{\psi(0)}$ the dressed state
\eqref{StatoCorretto} with a different transition frequency
$\omega_0'$, and also in this case a time evolution is expected. In
all the three cases, the evolution is based on the Hamiltonian
(\ref{IntHam}), according to which the atom has, for $t>0$, a
transition frequency $\omega_0$ (while the transition frequency is
$\omega_0'$ for $t<0$) . The results obtained in all three different
cases can be cast in the following compact form
\begin{widetext}\begin{equation}\label{3DeltaE}\begin{split}\Delta E^{(2)}_d(d)&=\lim_{m\to1}D_m\Bigl[\int_0^{+\infty}dx\frac{\sin(mx)}{x+x_0}\Bigr]\\
\Delta E^{(2)}_b(d,t)&=\lim_{m\to1}D_m\Bigl[\int_0^{+\infty}dx\frac{\sin(mx)}{x+x_0}\Bigl(1-\cos[a(x+x_0)]\Bigr)\Bigr]\\
\Delta E^{(2)}_p(d,t)&=\lim_{m\to1}D_m\Bigl[\int_0^{+\infty}dx\frac{\sin(mx)}{x+x_0}\Bigl(1-\cos[a(x+x_0)]\Bigr)+\int_0^{+\infty}\frac{\sin(mx)}{x+x'_0}\cos[a(x+x_0)]\Bigr]\\\end{split}\end{equation}\end{widetext}
where $a=ct/(2d)$, $x_0=2k_0d$, $x'_0=2k'_0d$, $x=2kd$ and $D_m$ is the differential operator
\begin{equation}\label{Diff}D_m=-\frac{\mu^2}{12\pi d^3}\Bigl[2-2\frac{\partial}{\partial m}+\frac{\partial^2}{\partial m^2}\Bigr].\end{equation}
The three interaction energies $\Delta E^{(2)}_d(d)$, $\Delta
E^{(2)}_b(d,t)$ and $\Delta E^{(2)}_p(d,t)$ in \eqref{3DeltaE} are,
respectively, for the fully dressed state, the bare state, and the
partially dressed state cases. The second and third interaction
energies reduce, as expected, to the first one for large values of
$a$ (that is of $t$). Moreover, the third one coincides with the
static expression for $k'_0=k_0$, since in this case the initial
state is the dressed ground state and then we do not expect any time
evolution.

The integrals appearing in Eq. \eqref{3DeltaE} can be calculated
analytically and expressed in terms of the sine and cosine integral
functions $\Si(x)$ and $\Ci(x)$ \cite{Abramowitz72}. The first
and the third integrals yield respectively
\begin{widetext}
\begin{equation}\label{3Int}\begin{split}
&\int_0^{+\infty}dx\,\frac{\sin(mx)}{x+x_0}=\Ci(mx_0)\sin(mx_0)+\frac{1}{2}\cos(mx_0)(\pi-2\Si(mx_0))\\
&\int_0^{+\infty}dx\,\frac{\sin(mx)}{x+x'_0}\cos[a(x+x_0)]=\frac{1}{4}\Bigl[-2\Ci[(a+m)x'_0]\sin[a(x_0-x'_0)-mx'_0]+2\Ci[l(a-m)x'_0]\sin[a(x_0-x'_0)+mx'_0]\\
&+\cos[a(x_0-x'_0)+mx'_0](-l\pi+2\Si[(a-m)x'_0])+\cos[a(x_0-x'_0)-mx'_0](\pi-2\Si[(a+m)x'_0])\Bigr]\end{split}\end{equation}
\end{widetext}
where $l=-1$ for $a<1$ $(t<2d/c)$ and $l=1$ for $a>1$ $(t>2d/c)$.
The integral in the second line of Eq. \eqref{3DeltaE} can be
obtained by just taking $k_0'=k_0$ in the second integral of Eq. \eqref{3Int}. Applying
the differential operator \eqref{Diff} and finally taking $m=1$, we
get the analytic expression of the interaction energy, from which
the atom-wall Casimir-Polder force can be obtained as the opposite
of the derivative with respect to the distance $d$. These
expressions are lengthy and not particularly enlightening from a
physical point of view and thus will be not reported here
explicitly.

\begin{figure}[h]\centering
\includegraphics[height=5cm]{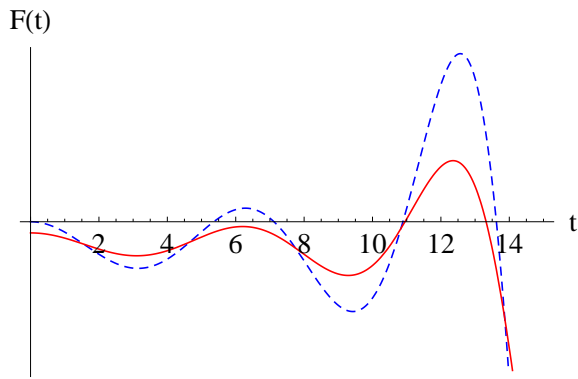}
\caption{(color online). Time evolution of the atom-wall force for $t<2d$ ($c=1$). We have chosen $k_0=1$, $k'_0=2$ and $d=10$, so that the back-reaction time is $t=20$. The red (solid) line corresponds to the case of an initial partially dressed state, while the blue (dashed) line is for the case of an initially bare ground state. Time is in units of $d/c$. The force is in arbitrary units.}
\label{FigPartially1}\end{figure}

\begin{figure}[h]\centering
\includegraphics[height=5cm]{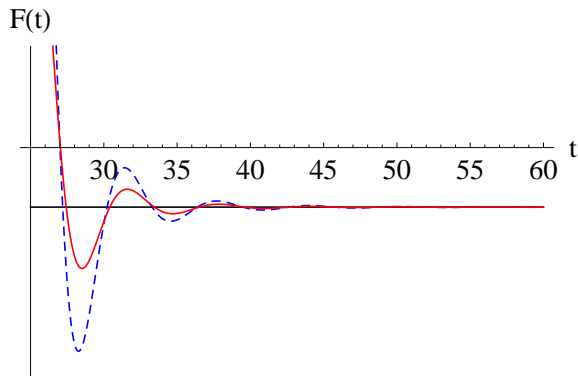}
\caption{(color online). Time evolution of the atom-wall force for $t>2d$, with the same values and units of Fig. \ref{FigPartially1}.}
\label{FigPartially2}\end{figure}

One main point of this paper is the comparison between the time
evolution of the force for the cases of an initial partially
dressed state and an initial bare ground state, the latter already
obtained in \cite{VasilePRA08}. In Figs. \ref{FigPartially1} and
\ref{FigPartially2} we give the plots of the time evolution of the
force for a bare ground state from \cite{VasilePRA08} (dashed blue
lines) and for a partially dressed state as obtained from $\Delta
E^{(2)}_p(d,t)$ in the third line of \eqref{3DeltaE} (solid red
lines). In both plots we take units such that $c=1$, and also
$k_0=1$ and $k_0'=2$; the atom is placed at a position such that
$k_0d=10$. The difference between the values used for $k_0$ and
$k_0'$ is quite large, and it has be chosen in such a way just for
the convenience of making more evident the qualitative different
features obtained in the two cases considered. Fig.
\ref{FigPartially1} refers to the region $a<1$, that is before a
light signal leaving the atom at $t=0$ reaches to the wall and
comes back, while Fig. \ref{FigPartially2} is for $a>1$. On the
light cone instead ($a=1$) the force diverges: the physical
meaning of this divergence, related to the well-known divergences
of source fields and to the dipole approximation, has been already discussed in
\cite{VasilePRA08}.

A first difference between the two cases (i.e. bare initial state
and partially dressed initial state) is that the initial ($t=0$)
value of the interaction energy for a partially dressed state is not
zero. This happens because, when the system is in a partially
dressed state at $t=0$, atom and field already \emph{see} each
other. It is interesting to analyze the time evolution towards the
asymptotic regime (i.e. for $a>1$). We see, as expected, that the
choice of a partially dressed initial state leads faster to the
asymptotic value of the force, exhibiting nevertheless a similar
oscillatory behavior around the asymptotic value. It is worth
stressing that the asymptotic value of the force, when the atom
becomes fully dressed, is the same in the two cases. This supports
the hypothesis that in many aspects the dynamics towards the fully
dressed state is indeed an irreversible process, with an equilibrium
state independent from the initial state \cite{PassanteOC93}.
The fact that different initial states, in general having different energies, lead for large times
to the same atom-wall potential energy is not in contradiction with the energy-conserving
unitary evolution of our system. The reason is that during the dynamical self-dressing
of the atom, a spike of radiation propagates on the light cone from the atom and asymptotically in time it carries away part of the energy
of the system to an infinite distance from the atom (see \cite{PassantePhysLettA03,CompagnoPRA88} for more detail). This energy is different for the cases considered (initially bare and partially dressed states), but it does not affect the large-time atom-wall interaction energy which
is related to the field fluctuations at the atomic position; the latter at large times occurs to be the same in the cases considered.

An important point is that, similarly to what found in
\cite{VasilePRA08} in the idealized case of an initial bare state,
also in the more realistic case of an initial partially dressed
atom, the force shows oscillations in time with negative
(attractive) and positive (repulsive) values. This oscillation of
the dynamical Casimir-Polder between an attractive and a repulsive
character, in the case of the partially dressed atom can in
principle be observed in the laboratory.

\begin{figure}[h]\centering
\includegraphics[height=5cm]{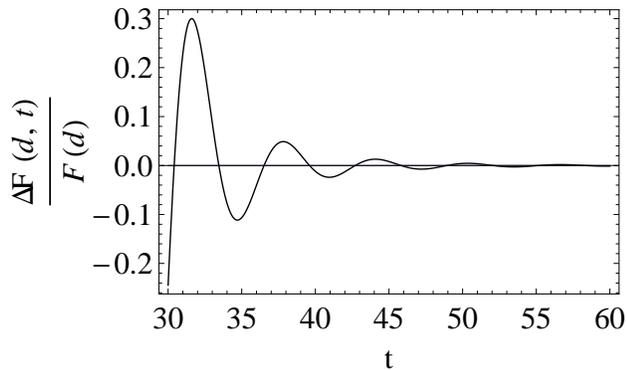}
\caption{Time evolution of the relative difference between the force
for an initially partially dressed state and the stationary value
for $t\to+\infty$. We have chosen $k_0=1$, $k'_0=2$ and $d=10$ (and
$c=1$). Time is in units of $d/c$.}
\label{RelDiff}\end{figure}

It is also significant to consider the evolution in time of the
relative difference between the dynamical force we computed and its
static value for $t\to+\infty$. We are thus going to consider the
quantity
\begin{equation}\frac{\Delta F(d,t)}{F(d)}=\frac{F_p(d,t)-F_d(d)}{F_d(d)}\end{equation}
with the same notations of Eq. \eqref{3DeltaE}. Fig. \ref{RelDiff} represents a plot of  $\Delta F(d,t)/F(d)$ with the same parameters as in Figs. \ref{FigPartially1} and \ref{FigPartially2}.
As expected, this relative
force difference oscillates in time and approaches zero for
$t\to+\infty$. In the next Section we shall discuss the orders of
magnitude of the physical parameters involved in the problem, as
well as observability of this new effect, that is the time-dependent
atom-wall Casimir-Polder force and its oscillatory behavior from an
attractive to a repulsive character.

\section{Discussion on the results}\label{Sec:4}

An essential point of our proposal outlined in the previous Section
for generating an atomic partially dressed state is to produce an
abrupt change of the atom's transition frequency from $\omega_0'$ to
$\omega_0$. In the present Section we shall propose a possible
method to realize this change and discuss the order of magnitude of
the relevant parameters involved. A possible technique to produce a
change of the atomic frequency is to place the atom at $t=0$ in a
uniform electric field of amplitude $E_0$, as first suggested in
\cite{PassantePhysLettA03,PassantePhysLettA96}. In this case,
assuming that the \emph{old} ($t<0$) free Hamiltonian of the system
is
\begin{equation}H'_0=\hbar\omega_0'\Sz+\sumkj\hbar\omegak\ackj\akj ,\end{equation}
the new ($t>0$) Hamiltonian is
\begin{equation}H_0=\hbar\omega_0\Sz+\sumkj\hbar\omegak\ackj\akj\end{equation}
where the difference between the new and the old frequency is related to the amplitude $E_0$ of the electric field.

We now address the problem of the timescale of the \emph{switching
on} of the electric field, and in particular if it is compatible
with our hypothesis that the quantum state of the system remains
unchanged immediately after this process. A reliable estimate of a
typical atomic evolution time is its inverse transition frequency
$\tau = \omega_0^{-1}$. Thus our \emph{non-adiabatic} hypothesis
becomes reasonable if the time necessary to switch on the electric
field is small compared to $\omega_0^{-1}$. Nevertheless, taking for
example the case of an hydrogen atom in its ground state, we have
$\tau=\omega_0^{-1}\simeq10^{-15}\,$s which seems to be a quite
short time to drive the electric field from zero to a value of $E_0$
sufficiently high to make appreciable our dynamical effects. This
difficulty in the experimental realization of the model discussed in
this paper, and the consequent observation of the dynamical
Casimir-Polder force, could be overcome by considering a Rydberg
atom, which can typically have a transition frequency of some GHz.
In this case, switching on an electric field in times shorter than
$\tau \sim 10^{-9}$ s should not be an impossible task (see
\cite{Pelekanos95}) and our assumptions should be valid. Our
assumption of a stable atomic ground state should be also valid with
a very good approximation in this case because Rydberg states can be
long-lived atomic states. An alternative method to generate a
partially dressed atomic state could be a rapid change of some other
physical parameter of the atom significantly affecting its
interaction with the radiation field, for example its refractive
index. This could be obtained by an optical control such that
obtained in \cite{Abdumalikov10}.

\section{Conclusions}

We have considered the dynamical atom-wall Casimir-Polder force in a
quasi-static approach for an initially partially dressed atom, and
compared in detail the results obtained with the case of an
initially bare state. A model for realizing the partially dressed
atom, as well its limits, has been discussed. The time evolution of
the atom-wall force has been calculated, and we have shown that it
exhibits oscillations in time yielding to a oscillatory change of
the Casimir-Polder force from an attractive to a repulsive
character, and that asymptotically in time it settles to the value
obtained in the stationary case. Possibility of experimental
verification of our results has been also discussed.

\begin{acknowledgments}

The authors thank the ESF Research Network CASIMIR for financial support.
They also acknowledge partial financial support from Ministero
dell'Universit\`{a} e della Ricerca Scientifica e Tecnologica and by
Comitato Regionale di Ricerche Nucleari e di Struttura della
Materia.
\end{acknowledgments}

\end{document}